\documentclass[showpacs,pre,amsmath,twocolumn,amssymb,floatfix]{revtex4-1}

\usepackage{epsfig}
\usepackage{graphicx}
\usepackage{dcolumn}

\begin{document}

\title{Impact induced transition from damage to perforation}

 \author{Attia Batool}
 \author{Gerg\H o P\'al}
  \author{Ferenc Kun}

  \email{Corresponding author: ferenc.kun@science.unideb.hu}
  \affiliation{Department of Theoretical Physics, Doctoral School of Physics, 
Faculty of Science and Technology, University of Debrecen, P.O.\ Box 400, H-4002 Debrecen, Hungary}
 \affiliation{Institute for Nuclear Research, Hungarian Academy of Sciences 
  (Atomki), P.O. Box 51, H-4001 Debrecen, Hungary}

  \date{\today}
  
\begin{abstract}
We investigate the impact induced damage and fracture of a bar shaped specimen of heterogeneous
materials focusing on how the system approaches perforation as the impact energy is gradually 
increased.  
A simple model is constructed which represents the bar as two rigid blocks coupled by a 
breakable interface with disordered local strength. The bar is clamped at the two ends and 
the fracture process is initiated by an impactor hitting the bar in the middle.
Our calculations revealed that depending on the imparted energy, the system has two phases: 
at low impact energies the bar suffers
damage but keeps its integrity, while at sufficiently high energies, complete perforation occurs.
We demonstrate that the transition from damage to perforation occurs analogous 
to continuous phase transitions. Approaching the critical point from below, the intact fraction 
of the interface goes to zero, while the deformation rate of the bar diverges according 
to power laws as function of the distance from the critical energy. 
As the degree of disorder increases, further from the transition point the critical exponents agree 
with their zero disorder counterparts, however, close to the critical point a crossover 
occurs to a higher exponent. 

\end{abstract}

\maketitle
  
\section{Introduction}
Under a slowly increasing mechanical load materials typically undergo  damaging 
and suffer ultimate failure at a critical load. The degree of materials' disorder
has been found to play an essential role in the evolution of the fracture process \cite{herrmann_statistical_1990,alava_statistical_2006,book_chakrabarti_2015}:
at high disorder cracks nucleate already at relatively low load levels resulting in a 
gradual accumulation of damage as the load increases so that global failure occurs 
as the culmination of damaging. This stable cracking is accompanied by the emission of
crackling noise \cite{rosti_crackling_2009} which proved to have a scale free 
statistics with a high robustness against materials' details \cite{garcimar_statistical_1997,
santucci_sub-critical_2004,meinders_scaling_2008,deschanel_experimental_2009,salje_crackling_2014}. 
The evolution of the fracture process 
has also been found to obey time-to-failure power laws when approaching failure 
addressing the possibility of forecasting the imminent catastrophic event
\cite{sornette_predictability_2002,nataf_predicting_2014,salje_main_minecollapse_2017,
kadar_record_2020}.  
In the opposite limit of low disorder, ultimate failure is preceded only by a small 
amount of damage, an unstable crack emerges which gives rise to an abrupt failure
\cite{book_chakrabarti_2015}.

Experimental and theoretical studies have revealed that fracture of disordered 
materials shows interesting analogies to phase transitions 
and critical phenomena \cite{zapperi_first-order_1997,biswas_lls_2017,book_chakrabarti_2015}. 
In particular, loaded solids can be considered to be in a meta-stable state \cite{sethna_prl1996}
so that the point of failure has been interpreted as a nucleation process in a first 
order phase transition near a spinodal 
\cite{zapperi_first-order_1997,zapperi_analysis_1999,kun_damage_2000}. 
Other studies suggested that the 
transition from damage to fracture of highly disordered materials is analogous 
to continuous phase transitions due to the universal scaling laws emerging 
in the system near the critical point \cite{sornette_scaling_1998,moreno_fracture_2000,
caldarelli_self_organization_1996,bonamy_crackling_prl2008,roy_phasetrans_pre2019}.
Recently, it has been clarified that tuning the amount of materials' disorder
a transition occurs from brittle to quasi-brittle behaviors which proved to be
continuous for long range stress redistribution \cite{picallo_brittle_2010,udi_epl_2011,danku_disorder_2016,
manna_PhysRevE.91.032103,ramos_prl_2013,ray_epl_2015}. 

Studies on the phase transition nature of fracture phenomena have mostly been focused on 
uni axial quasi-static loading conditions. However, both the time evolution and 
the final outcome of fracture processes also depend on how the load is
applied on the specimen. For instance, in the usual Charpy impact test of dynamic fracture 
\cite{siewert_pendulum_2000,francois_charpy_2002,kun_structure_2004,danku_apl_2015}, 
the specimen is clamped at the ends and a hammer attached to the arm of a pendulum 
hits it in the middle resulting in a dynamic three point bending. 
Under such boundary and loading conditions, the damage localizes to a relatively thin
layer of the specimen giving rise to a single growing crack.
Depending on the energy of the hit,
the crack can either terminate and the specimen suffers only partial
failure, or it runs to the opposite boundary breaking the specimen into two pieces.
This impact induced transition from damage to complete perforation is driven by the interplay 
of materials disorder and the inhomogeneous stress field emerging due to the 
bending loading. 
In the present paper we study this transition as the energy of impact 
is varied. Based on a simple stochastic interface model of bending induced breaking
of bar shaped specimens, we demonstrate that the transition is analogous to continuous 
phase transitions. Approaching the critical energy of perforation the fraction of intact
cross section goes to zero, while the deformation rate diverges both according to power laws 
as function of the distance from the critical point. We show that the critical exponents 
exhibit a crossover when the degree of disorder is varied, 
however, the transition remains continuous even in the limit of zero disorder.

\section{Stochastic interface model}
\begin{figure}
\epsfig{bbllx=40,bblly=430,bburx=540,bbury=750, file=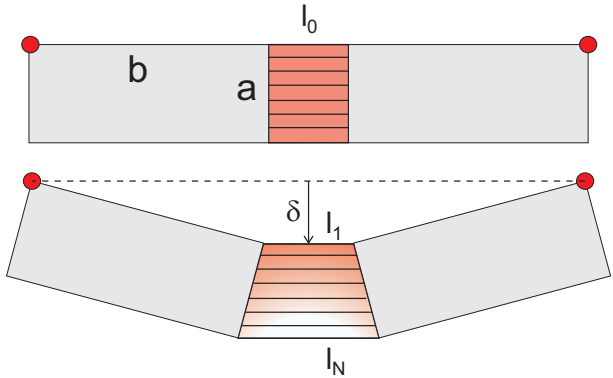, width=8.3cm}
\caption{Model of the specimen under three point bending. The specimen is composed of two 
rigid blocks of extensions $a$ and $b$ coupled by an elastic interface. The interface is discretized
in terms of breakable fibers with stochastic strength. The deformation of the bar 
due to impact is characterized by the deflection $\delta$ of its middle point. The impact 
loading results in a linear deformation profile along the interface, where the top and bottom
fibers have the smallest $l_1$ and largest $l_N$ length, respectively. 
\label{fig:setup}}
\end{figure}

To investigate the damage-perforation transition in impact loading processes, 
we use a simple model of the three-point bending setup of bar shape specimens.
In the model the specimen is represented by two rigid blocks of side length $a$ and $b$,
which are glued together by a thin deformable interface of width $l_0\ll b$.
Clamping of the ends of the specimen is 
ensured by fixing the upper outer corners of the blocks around which
they can perform rigid rotation. External loading is exerted by an impactor which 
hits the bar in the middle with a velocity pointing downward as it is illustrated 
in Fig.\ \ref{fig:setup}.
As a consequence, the bar gets deflected in such a way that the entire deformation
is accommodated by the elastic interface between the blocks.
In order to capture fracturing of the bar, the interface is discretized by means of a 
bundle of parallel fibers of number $N$ and length $l_0$, placed equidistantly between 
the blocks. The fibers do not have a bending rigidity, hence, they suffer only stretching when 
the bar is deflected. Fibers are assumed to have a linearly elastic behavior 
up to a threshold deformation $\varepsilon_{c}^i, i=1,\ldots , N$ where they break irreversibly. 
Disorder of the material is represented such that individual fibers are characterized 
by an identical Young modulus $Y$, however, their breaking threshold is a random variable sampled
from a probability density function $p(\varepsilon_c)$. 

Due to the rigidity of the blocks the deformation of the specimen can be represented by a single 
variable $\delta$, which is the deflection of the middle point of the bar. For illustration
of the geometrical setup and the loading condition see Fig.\ \ref{fig:setup}.
At a finite value of $\delta>0$ the interface fibers suffer elongation $\Delta l$ 
which increases from top to bottom resulting in the opening of the interface. 
Based on the geometrical setup of Fig.\ \ref{fig:setup}, the actual length of fibers $l_i$ 
can be expressed as
\begin{equation}
 l_i = l_1+2\delta\frac{a}{b}\frac{i-1}{N-1}, \qquad i=1,\ldots,N
\end{equation}
where index $i$ identifies the position of fibers starting from the top of the interface.
The length of the first fiber $l_1$ reads as $l_1=l_0+2(b-\sqrt{b^2-\delta^2})$. 
Finally, the local elongation $\Delta l_i=l_i-l_0$ of fibers can be cast into the form
\begin{equation}
\Delta l_i = 2b-2\sqrt{b^2-\delta^2}+2\delta\frac{a}{b}\frac{i-1}{N-1},
\label{eq:gradient}
\end{equation}
which yields a linear deformation profile $\varepsilon_i=\Delta l_i/l_0$ along the interface.

Fracture is initiated by a collision with a body of mass $m$ which hits the 
bar in the middle with an initial velocity $v_0$. 
For simplicity, we assume that the mass of the specimen is negligible compared to the impactor,
furthermore, the impactor and the bar stay in contact during the entire fracture process.
Hence, the initial kinetic energy $E_0=1/2mv_0^2$ imparted to the system
will be partly transformed into the elastic energy $E_{el}$ of the elongated fibers 
and it gets partly dissipated $E_{dis}$ by the breaking fibers.
The energy balance of the collision process can be written in the form
\begin{equation}
E_0 = E_{k}(\delta (t)) + E_{el}(\delta (t)) + E_{dis}(\delta (t)),
\label{eq:balance}
\end{equation}
at any time $t$ as the system evolves.
On the right hand side of Eq.\ (\ref{eq:balance}) the first term $E_k$ 
denotes the kinetic energy of the impactor
\begin{equation}
E_k = \frac{1}{2}mv^2
\end{equation}
moving at velocity $v$, which can be expressed as the derivative of the deflection
$v=d\delta/dt$ of the specimen. 

As the deflection increases fibers gradually get deformed and eventually break when exceeding 
their local breaking threshold. The elastic energy $e_{el}$ stored by a single fiber of elongation 
$\Delta l$ can be obtained as $e_{el}=\left(\frac{ac}{2Nl_0}\right)Y\Delta l^2$, where the cross
sectional area $ac/N$ is assigned to the fiber with $c=1$ being the unit thickness of the sample.
Hence, the energy $E_{el}$ stored in the deformation of the entire bundle at deflection $\delta$ 
can be expressed analytically in terms of the disorder distribution of fibers as
\begin{equation}
E_{el}(\delta) = \frac{ac}{2Nl_0}Y\sum_{i=1}^N\left[1-P(\varepsilon_i(\delta))\right]
\Delta l_i(\delta)^2. 
\label{eq:e_el}
\end{equation}
Here $P(x)$ represents the cumulative distribution of breaking thresholds so that 
the term $\left[1-P(\varepsilon_i(\delta))\right]$ is the probability that the fiber at location 
$i$ along the interface remained intact at the deflection $\delta$. 
The energy $E_{dis}$ dissipated by the broken fibers can be obtained as
\begin{equation}
E_{dis} = \frac{ac}{2Nl_0}Y\sum_{i=1}^N\int_0^{\Delta l_i(\delta)} p(x)x^2dx,
\label{eq:e_dis}
\end{equation}
assuming that the elastic energy stored in a fiber at the instant of its breaking 
is consumed to create the corresponding crack surface.
\begin{figure}
\begin{center}
\epsfig{bbllx=30,bblly=20,bburx=380,bbury=330, file=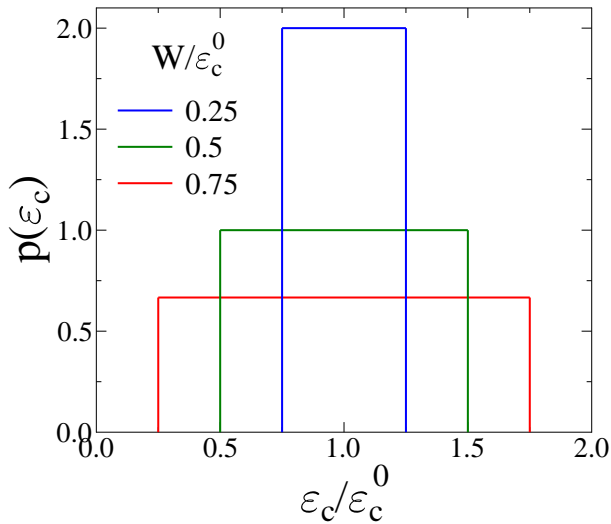, width=8.3cm}
  \caption{Probability distribution $p(\varepsilon_c)$ of the breaking thresholds 
  for three different values 
  of the width $W/\varepsilon_c^0=0.25, 0.5, 0.75$. The average strength of fibers 
  $\varepsilon_c^0$ is fixed, however, the degree of disorder can be tuned by varying the width
  $W$ of the distribution.
   \label{fig:probdens}}
\end{center}
\end{figure}

In order to study the effect of the degree of materials' disorder on the damage - perforation
transition, for the breaking thresholds we introduce a uniform distribution of the form
\begin{equation}
p(\varepsilon_c) = 1/(2W) \ \ \ \mbox{for} \ \ \ \varepsilon_c^0-W\leq\varepsilon_c\leq\varepsilon_c^0+W.
\label{eq:probdens}
\end{equation}
The distribution $p(\varepsilon_c)$ is centered on $\varepsilon_c^0$ which denotes 
the average strength of fibers. The value of $\varepsilon_c^0$ is fixed in all the calculations, 
however, the degree of disorder of the system can be controlled by 
varying the width of the distribution $W$ in the range $0\leq W \leq \varepsilon_c^0$. 
The disorder distribution $p(\varepsilon_c)$ is illustrated in Fig.\ 
\ref{fig:probdens} for several values of $W$. 
\begin{figure}
\epsfig{bbllx=30,bblly=20,bburx=380,bbury=330, file=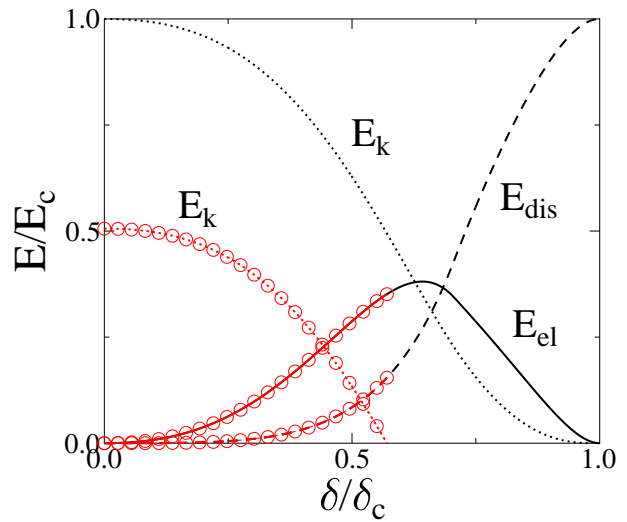, width=8.3cm}
\caption{The evolution of the energies during an impact process at the
critical impact energy $E_0 = E_c$ as the function
of deflection $\delta$ (black lines). In the final state $E_c$ is completely dissipated by fiber
breaking, hence, $E_{el}(\delta_c) = 0$ and $E_{dis}(\delta_c) = E_c$. Simulations were
performed for an interface of $N=10^6$ fibers with $W=\varepsilon_c^0$ 
for the threshold distribution Eq.\ (\ref{eq:probdens}). The red symbols highlight the case 
of a sub-critical impact at $E_0\approx E_c/2$.
\label{fig:ener}}
\end{figure}

As the bar gets deflected during the impact process, a linear deformation profile builds up 
along the interface according to Eq.\ (\ref{eq:gradient}). In the absence of disorder $W=0$ 
all breaking thresholds
are the same $\varepsilon_c^i=\varepsilon_c^0$ $(i=1,\ldots , N)$ so that the fiber at the bottom
of the interface $(i=N)$ breaks first. As a consequence, a crack starts and advances upwards 
until the last fiber $(i=1)$ breaks.
However, in the presence of disorder $W>0$, the local strength $\varepsilon_c^i$ and strain 
$\varepsilon_i$ together determine the order of breaking so that the breaking sequence 
of fibers becomes random along the interface. 
Inverting Eq.\ (\ref{eq:gradient}) we can determine the deflection values $\delta_c^i$ 
where the individual fibers break
\begin{equation}
\delta_c^i = \delta(i,\varepsilon_c^i), \qquad i=1, \ldots, N.
\label{eq:defthres}
\end{equation}
Note that the deflection thresholds $\delta_c^i$ depend on both the position $i$ and 
strength $\varepsilon_c^i$ of the fiber, which are independent variables.
As the bar gets gradually deflected, fibers break in the increasing
order of their deflection thresholds $\delta_c^i$, $(i=1,\ldots , N)$ resulting in a 
random sequence of breaking events along the interface in the linear deformation profile.

Computer simulation of the impact process is performed in the following way:
in the initial state random threshold values $\varepsilon_c^i$ (i=1,\ldots, N) are 
assigned to each fiber from the distribution Eq.\ (\ref{eq:probdens}). 
The deflection thresholds $\delta_c^i$ are determined from Eq.\ (\ref{eq:defthres}), 
which are then sorted into ascending order. The dissipated $E_{dis}$ and elastic $E_{el}$ 
energies are then calculated from the discrete form of Eqs.\ (\ref{eq:e_el},\ref{eq:e_dis}) 
as function of $\delta$.
To analyze the damage-perforation transition, and to quantify the role of disorder,
numerical calculations were performed varying the number of fibers in
a broad range $N=10^3 - 10^7$ at several values of the disorder parameter 
$W/\varepsilon_c^0\in [0,1]$. The geometrical layout of the specimen was fixed to
$a=1$, $b/a=2.5$, and $l_0/a=0.02$.

\section{Energetics of the loading process}
When the impactor hits the bar with an initial kinetic energy $E_0$, the bar gets 
deflected and the interface fibers start to break. 
If the input energy is low the damage process stops at a maximum deflection $\delta_m$ 
where the interface suffers only a partial breaking keeping the integrity of the sample.
When the maximum deflection $\delta_m$ is reached, the impactor stops, hence, the sum 
of the elastic and dissipated energies  must be equal to the initial energy $E_0$ of impact
\begin{equation}
E_0 = E_{el}(\delta_m)+E_{dis}(\delta_m). 
\label{eq:e0}
\end{equation}
Inserting Eqs.\ (\ref{eq:e_el},\ref{eq:e_dis}) into Eq.\ (\ref{eq:e0}) 
we can determine the initial impact energy $E_0$
needed to achieve a maximum deflection $\delta_m$. When $E_0$ is sufficiently high,
the damage process does not stop, all fibers break so that the specimen gets perforated.
This first occurs at the critical impact energy $E_c$, where deflection stops right at the 
breaking of the last fiber, with the largest critical deflection $\delta_c^i$, 
defining the critical deflection $\delta_c$ of the system.
It follows that the critical energy $E_c$ is equal to the total dissipated energy 
$E_{dis}(\delta_c)$ needed to break through the entire interface.
\begin{figure}
\epsfig{bbllx=30,bblly=20,bburx=380,bbury=330, file=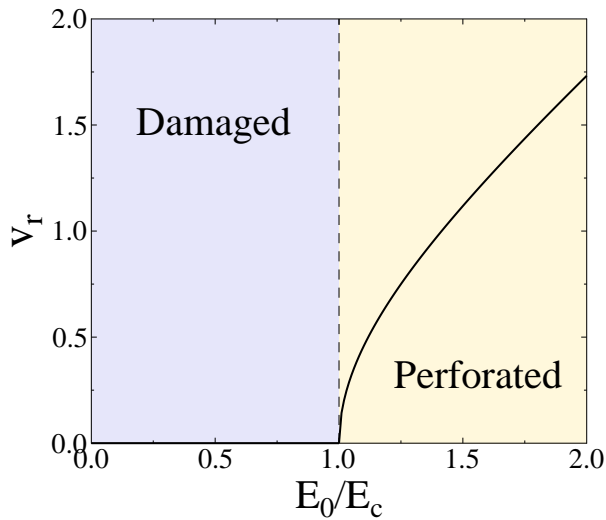, width=8.3cm}
\caption{The remaining velocity  $v_r$  of the impactor as a function of the imparted 
energy $E_0$. In the damage phase the impactor stops $v_r=0$ at a partial failure of the 
specimen, while beyond the perforation limit $E_0>E_c$ it keeps moving after the specimen broke 
into two pieces. The impactor is assumed to have a unit mass. 
\label{fig:ballistic}}
\end{figure}

The evolution of the energies stored in deformation $E_{el}(\delta)$ and dissipated by 
fiber breaking $E_{dis}(\delta)$, furthermore, the remaining kinetic energy $E_k(\delta)$
of the impactor are illustrated in Fig.\ \ref{fig:ener} 
as function of the deflection $\delta$ of the specimen at the critical energy $E_0=E_c$
for a system containing $N^6$ fibers at the highest disorder $W/\varepsilon_0^c=1$. 
The remaining kinetic energy of the impactor $E_k(\delta)$ was obtained as 
\begin{equation}
E_k(\delta) = E_0-\left[E_{el}(\delta)+E_{dis}(\delta)\right],
\end{equation}
where the impact energy $E_0$ is determined from Eq.\ (\ref{eq:e0}) at 
$\delta_m=\delta_c$
Since the disorder distribution in the example extends down to zero strength values, breaking
starts already at very small deflection. It can be observed in Fig.\ \ref{fig:ener} 
that the dissipated 
energy $E_{dis}$ monotonically increases, while the elastic energy $E_{el}$
has a maximum at a well defined deflection value. The remaining kinetic energy $E_k$
monotonically decreases reaching zero at the critical deflection.
The figure also highlights the energetics of a sub-critical impact $E_0\approx E_c/2$ 
where the bar suffers damage in the form of fiber breaking, 
but does not perforate. Since the elastic and dissipated energies
are fully determined by the deflection of the bar $\delta$, these curves coincide
with their counterparts obtained at the critical input energy $E_0=E_c$. 
However, at the maximum deflection $\delta_m$, indicated by the end of the red curves, 
the kinetic energy is zero since the impactor stops.
\begin{figure}
\epsfig{bbllx=30,bblly=20,bburx=380,bbury=330, file=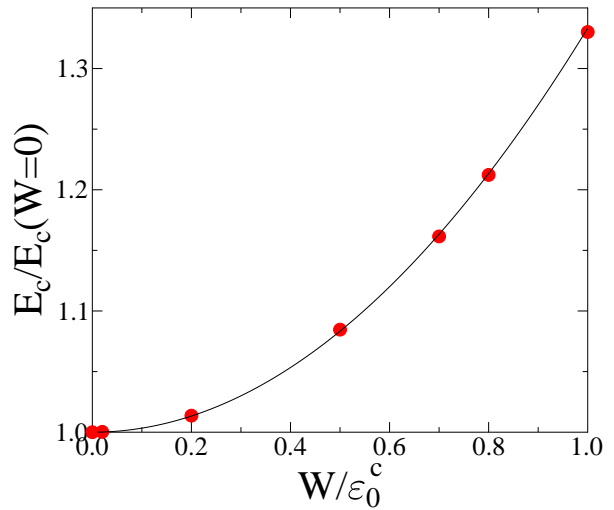, width=8.3cm}
\caption{The critical energy $E_c$ of perforation 
as a function of the amount of disorder $W/\varepsilon_0^c$. 
The continuous line represents the analytical 
solution Eq.\ (\ref{eq:critener}), while the symbols stand for numerical measurements.
$E_c$ is divided by the critical energy of the zero disorder case.
\label{fig:critener}}
\end{figure}

In the super-critical phase $E_0/E_c>1$ the specimen perforates, and hence, the impactor 
does not stop. The curves of the elastic and dissipated energies still coincide with the ones
obtained at the critical point $E_0/E_c=1$, however, the kinetic energy does not decrease 
to zero. Instead, the impactor continues moving with the remaining energy $E_{k}$ attained at 
the critical deflection $\delta_c$. This super-critical regime is 
illustrated by the so-called ballistic diagram of the system in Fig.\ 
\ref{fig:ballistic} where the remaining velocity $v_r$ of the impactor is plotted as a function 
of the input energy.
The value of $v_r$ was determined as
\begin{equation}
v_r = \sqrt{\frac{2}{m}\left(E_0-E_c\right)},  \qquad \mbox{for} \qquad E_0\geq E_c.
\end{equation}
The two phases of the impact process are also highlighted 
in Fig.\ \ref{fig:ballistic}: in the damaged phase, below the critical impact energy, 
the impactor stops $v_r=0$ at the maximum deflection $\delta_m$ reached, 
while the perforated phase is characterized by a finite value of the remaining velocity $v_r>0$
since after breaking the bar the impactor retains a finite fraction of its initial energy.

\section{Effect of disorder and of the finite number of fibers on the critical point}
It is a crucial question how materials' disorder affects the impact induced fracture 
of specimens. 
\begin{figure}
\epsfig{bbllx=20,bblly=100,bburx=730,bbury=1010, file=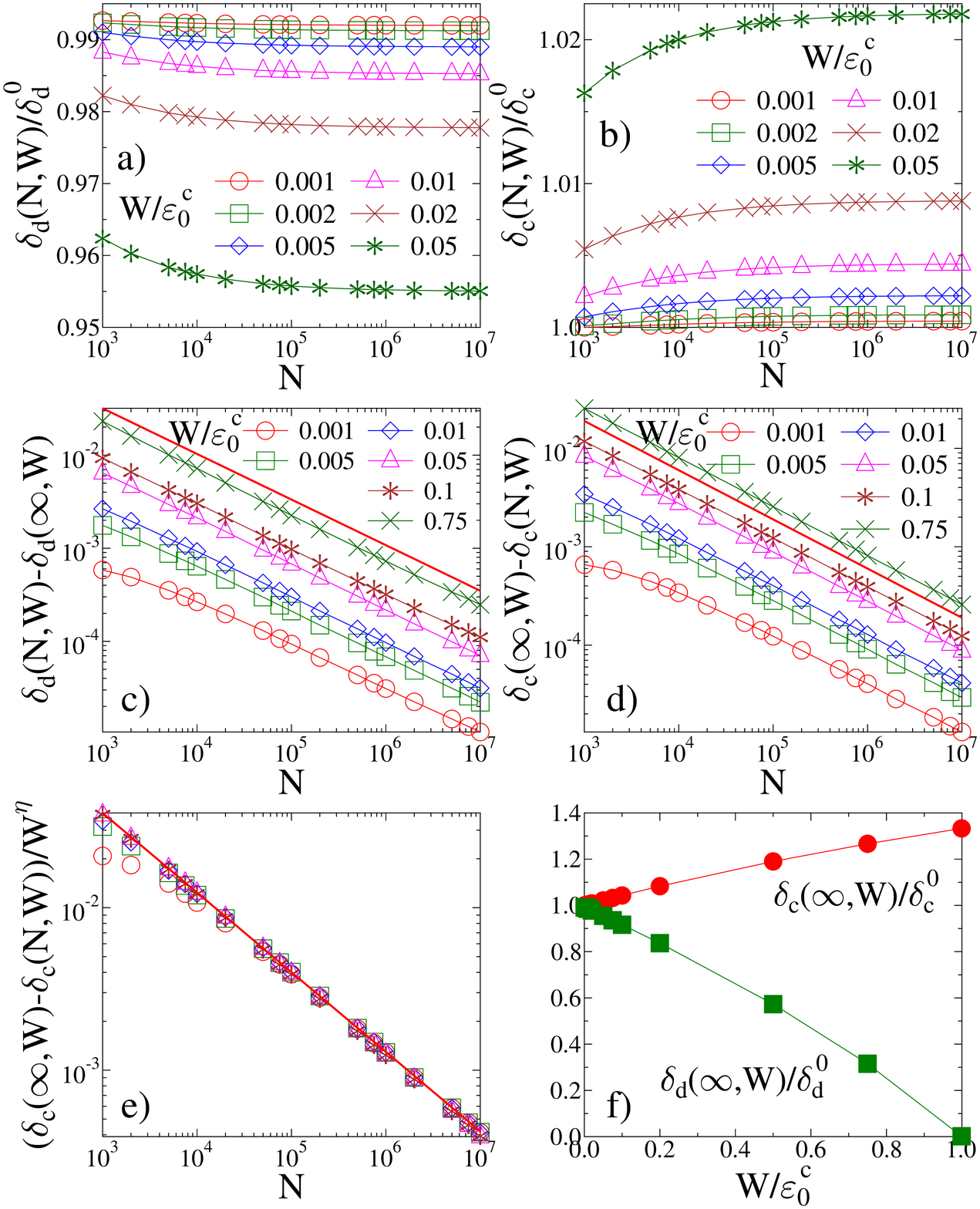, width=8.3cm}
\caption{The effect of the number of fibers $N$ and the degree of disorder $W$ on the 
damage threshold $\delta_d$ and on the critical deflection $\delta_c$ of the system.
The damage threshold $\delta_d(N,W)$ $(a)$ and the critical deflection $\delta_c(N,W)$ $(b)$
are presented as function of $N$ for several values of $W$. Both quantities converge to 
asymptotic values $\delta_d(\infty,W)$ and $\delta_c(\infty,W)$ in the limit $N\to\infty$.
The deviation from the asymptotic value has a universal power law dependence on the number 
of fibers in $(c)$ and $(d)$. The bold straight lines represent power laws of exponent $-1/2$.
Rescaling the deviations by an appropriate power of $W$,
the curves obtained at different disorders can be collapsed on the top of each other. 
This scaling is demonstrated for $\delta_c(N,W)$ in $(e)$. The asymptotic values of the damage 
threshold and critical deflection as function of $W$ $(f)$.
\label{fig:sizeeffect}}
\end{figure}
Based on the disorder distribution Eq.\ (\ref{eq:probdens}) 
and on the expression of the dissipated energy Eq.\ (\ref{eq:e_dis})
the critical energy $E_c$ can be easily obtained. Since the energy 
dissipated by the breaking of a single fiber depends only on its breaking threshold $\varepsilon_c$
but not on the corresponding deflection $\delta_c$, the integral of Eq.\ (\ref{eq:e_dis}) 
simplifies to integration over the range of the $\varepsilon_c$ values.  
Hence, the disorder dependence of the critical energy $E_c$ can be cast into the form
\begin{equation}
E_c =\frac{ac}{12Wl_0}\left[\left(\varepsilon_0^c+W\right)^3 - \left(\varepsilon_0^c-W\right)^3\right],
\label{eq:critener}
\end{equation}
It follows that $E_c$ monotonically increases from $E_c(W=0)=(ac/2l_0)(\varepsilon_c^0)^2$ at zero disorder 
to $E_c(W=\varepsilon_c^0)=(2ac/3l_0)(\varepsilon_c^0)^2$ at the highest one. 
It can be observed in Fig.\ \ref{fig:critener} that the results of numerical measurements 
agree very well with the analytical expression Eq.\ (\ref{eq:critener}) of the critical
energy. 

Since the critical energy $E_c$ is an integrated quantity of the entire sample it has 
only very low fluctuations when calculated from single samples using a finite number $N$
of fibers to discretize the interface. However, the deflection values where the damaging 
starts $\delta_d$ and perforation occurs $\delta_c$ do have a strong dependence 
both on the number of fibers $N$ and on the degree of disorder $W$. In order to quantify 
these effects, we performed computer simulations of the impact process varying $N$ over four 
orders of magnitudes at several values of $W$. For a single sample the damage threshold 
$\delta_d$ and the critical deflection $\delta_c$ were identified as the deflection values 
where the first and last fiber breaking occur, respectively. Then these values were averaged 
over $1000$ samples at each parameter set.

It can be observed in Fig.\ \ref{fig:sizeeffect}$(a,b)$ that both the damage threshold $\delta_d$
and the critical deflection $\delta_c$ depend on the number of fibers $N$ used to discretize 
the interface. However, $\delta_d$ decreases, while $\delta_c$ increases converging 
to well defined asymptotic
values  $\delta_d(\infty,W)$ and $\delta_c(\infty,W)$ in the $N\to\infty$ limit. 
Note that both quantities are normalized by their zero 
disorder counterparts
\begin{eqnarray}
\delta_d^0 &=& a\left[\sqrt{1+\frac{bl_0\varepsilon_c^0}{a^2}}-1\right], \label{eq:dam_crit_zer}\\
\delta_c^0 &=& \frac{l_0\varepsilon_c^0}{2}\sqrt{\frac{4b}{l_0\varepsilon_c^0}-1}, \label{eq:crit_crit_zer}
\end{eqnarray}
which do not depend on $N$. It can be seen that increasing the amount of disorder 
$W$, the asymptotic values $\delta_d(\infty,W)$ and $\delta_c(\infty,W)$ 
more and more deviate from the zero 
disorder values. Figures \ref{fig:sizeeffect}$(c,d)$ demonstrate that the convergence
of the damage threshold $\delta_d(N,W)$ and of the critical deflection 
$\delta_c(N,W)$ to the corresponding asymptotic values shows interesting universal 
features, i.e.\ at any value of $W>0$ the distances $\delta_d(N,W)-\delta_d(\infty,W)$ and 
$\delta_c(\infty,W)-\delta_c(N,W)$ tend to zero as a universal power law of the number 
of fibers $N$. The bold straight lines in the figures represent power laws of exponent $-1/2$.
The value of the exponent does not depend on the degree of disorder, however, the curves get 
shifted with respect to each other for increasing $W$.
Figure \ref{fig:sizeeffect}$(e)$ shows that rescaling the critical deflections $\delta_c$ 
with an 
appropriate power $\eta$ of the degree of disorder $W$, the curves obtained at different 
$W$ values can be collapsed on the top of each other. (The same systematics holds also 
for the damage threshold $\delta_d$, not shown in the figure.)
Based on the above numerical analysis we can cast the $N$ and $W$ dependence of the damage 
threshold and of the critical deflection into the following scaling forms
\begin{eqnarray}
\delta_d(N,W) &=& \delta_d(\infty,W) + AW^{\eta}N^{-\mu}, \label{eq:scaling_d} \\
\delta_c(N,W) &=& \delta_c(\infty,W) - BW^{\eta}N^{-\mu}, \label{eq:scaling}
\end{eqnarray}
where the scaling exponents have the values $\eta=1/2$ and $\mu=1/2$.

Eqs.\ (\ref{eq:scaling_d},\ref{eq:scaling}) highlight that the asymptotic values 
of the damage and perforation
thresholds both depend on the degree of disorder $W$. We determined  $\delta_d(\infty,W)$
and $\delta_c(\infty,W)$ in such a way that they were finely tuned to obtain 
the best straight lines in Figs.\ \ref{fig:sizeeffect}$(c,d)$ on a double logarithmic plot.
The final outcome is presented in Fig.\ \ref{fig:sizeeffect}$(f)$ where both the asymptotic 
damage threshold and critical deflection are normalized by their zero disorder counterparts
given by Eqs.\ (\ref{eq:dam_crit_zer},\ref{eq:crit_crit_zer}). 
It can be observed that for growing disorder the damaging
of the interface starts earlier while perforation is reached at a higher deflection if a 
sufficient amount of energy is exerted to the system.
An important outcome of the analysis is that for a sufficiently fine discretization
the results do not depend on the number of fibers.

In the following study of the perforation transition the number of fibers of the 
discretization was fixed to $N=10^6$.

\begin{figure}
\epsfig{bbllx=30,bblly=20,bburx=380,bbury=330, file=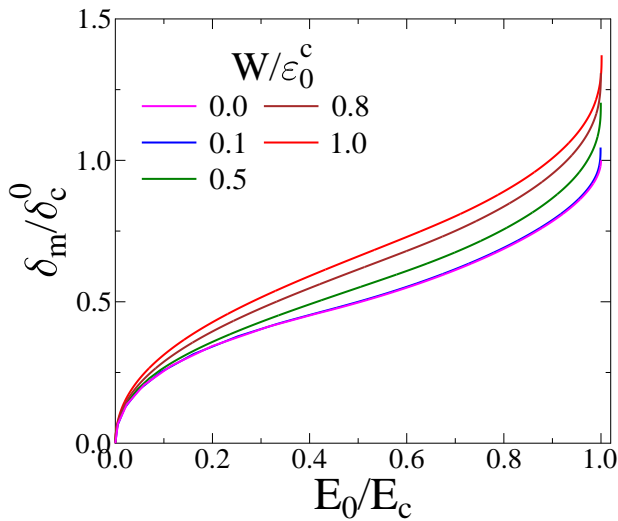, width=8.3cm}
\caption{Maximal deflection $\delta_m$ of the bar reached as a function of the impact
energy $E_0$ for several disorder strengths $W$. 
The value of $\delta_m$ is non-dimensionalized by division with the critical deflection
$\delta_c^0$ of the zero disorder case.
\label{fig:deflec}}
\end{figure}

\section{Approach to the critical point}
In order to understand how the system approaches the critical point 
of perforation, numerical calculations were performed varying the input energy $E_0$
below $E_c$ at several values of the degree of disorder $W$. 
Figure \ref{fig:deflec} presents the maximum deflection $\delta_m$ reached during 
the impact process as a function of the imparted energy $E_0$.
It can be observed that $\delta_m$ monotonically increases with $E_0$
up to the critical deformation $\delta_c=\delta_m$ attained at $E_0=E_c$. 
However, the deflection rate $\delta_m^{\prime}=d\delta_m/dE_0$ is not monotonous, i.e.\ 
it can be inferred from the curves that
starting from a relatively high value $\delta_m^{\prime}$ first decreases at low energies
reaching a finite minimum which is then followed by a growing deformation rate and seems
to diverge as
the critical energy is approached. Note that although the critical energy $E_c$ is an 
increasing function of the degree of disorder $W$ (see Fig.\ \ref{fig:critener}), i.e.\
higher energy is needed to break the sample, at the same fraction $E_0/E_c$ of $E_c$ 
the maximum deflection $\delta_m$ takes a larger value for higher $W$.

\begin{figure}
\epsfig{bbllx=30,bblly=20,bburx=380,bbury=330, file=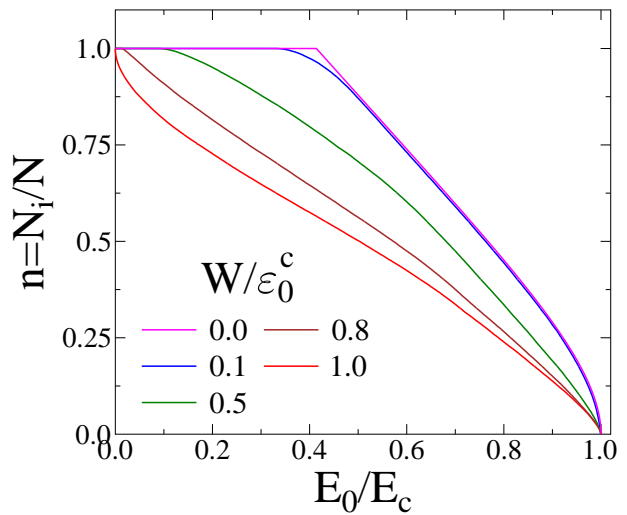, width=8.3cm}
\caption{Fraction of intact fibers $n=N_i/N$ as a function of the impact energy 
$E_0$ for several values of the degree of disorder $W$.
\label{fig:intfract}}
\end{figure}

Gradually increasing the energy of impact, the increasing final state deformation 
is accompanied by a growing damage of the 
interface. To characterize the damage state of the bar we determined the fraction of intact 
fibers $n=N_i/N$, where $N_i$ denotes the number of surviving fibers when the maximum deflection 
is reached at $E_0$. 
Figure \ref{fig:intfract} illustrates that $n$ is a monotonically decreasing function
of the imparted energy $E_0$ starting from one and converging to zero as the critical point 
of perforation $E_c$ is approached. Note that at lower disorder the $n(E)$ curves start with a
constant regime $n=1$, since the final state deflection has to surpass the damage threshold 
$\delta_m>\delta_d$ to initiate breaking of fibers.
The figure also demonstrates that 
at a given fraction of the critical energy $E_0/E_c$ the surviving load bearing cross section $n$
of the sample gets lower implying a higher damage $d=1-n$ when the disorder is higher.

In order to understand the nature of the transition from partial failure to perforation,
we analyzed in details how the system behaves in the vicinity of the critical point $E_c$.
It has been shown above that as the input energy increases, the critical point of perforation 
is approached 
through an acceleration of the deformation of the bar in such a way that $\delta_m^{\prime}$ 
diverges in the limit $E_0\to E_c$. It means that close to $E_c$ the system becomes more 
and more susceptible to the gradual increase of the impact energy responding with a rapidly
growing deflection. For a quantitative characterization of this susceptibility, 
we determined the deflection rate $\delta_m^{\prime}$ 
by numerical differentiation of the $\delta_m(E_0)$ curves for several disorder strengths $W$. 
\begin{figure}
\epsfig{bbllx=30,bblly=20,bburx=380,bbury=330, file=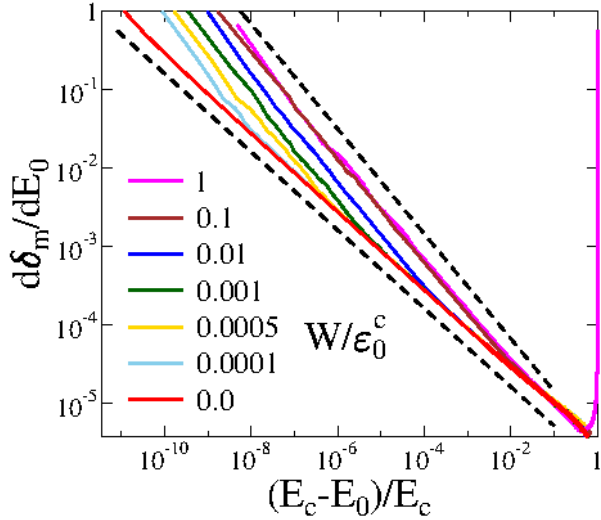, width=8.3cm}
\caption{Derivative of the maximal deflection $\delta_m$ with respect to the impact energy 
$E_0$ as a function of the relative distance from the critical point $(E_c-E_0)/E_c$ 
for several values of the degree of disorder $W$. The dashed lines represent power 
laws of exponent $1/2$ and $2/3$. As the disorder $W$ increases the crossover point between 
the two power law regimes shifts to the right.
\label{fig:deriv}}
\end{figure}
Figure \ref{fig:deriv} demonstrates that for zero disorder $W=0$ the deflection rate 
diverges as a power law of the distance from the critical point
\begin{equation}
\delta_m^{\prime} \sim (E_c-E_0)^{-\gamma},
\label{eq:critdefrate}
\end{equation}
where the exponent $\gamma$ has the value $\gamma=1/2$. Note that the quality of the
power law approximation is excellent, for a system of $N=10^6$ fibers a straight line 
is obtained on a double logarithmic plot over 8 orders of magnitude. 
It is important to emphasize that at finite disorder $W>0$ when the fibers have a stochastic 
variation of strength, the power law divergence prevails, however, a crossover occurs
between two regimes of different exponents: further from $E_c$ the value of the power law exponent
coincides with its zero disorder counterpart $\gamma=1/2$, however, close to the critical point
the deflection accelerates characterized by a higher exponent $\gamma=2/3$.

\begin{figure}
\epsfig{bbllx=30,bblly=20,bburx=380,bbury=330, file=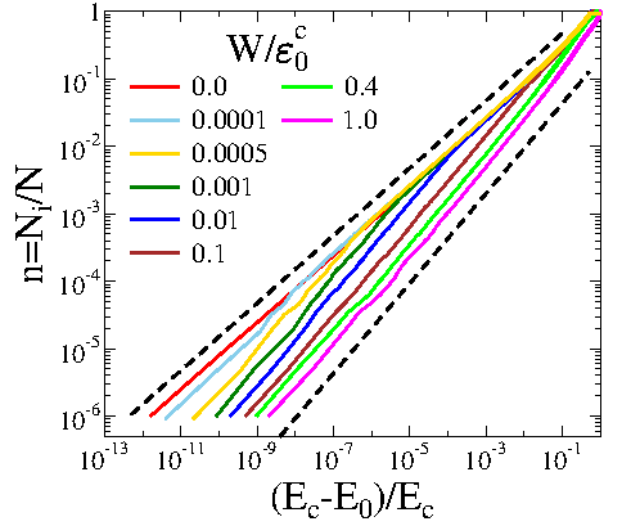, width=8.3cm}
\caption{The fraction of intact fibers as a function of the relative distance from the critical 
point $\Delta=(E_c-E_0)/E_c$ for several values of the degree of disorder $W$. 
The dashed lines represent power laws of exponent $1/2$ and $2/3$. 
\label{fig:intfrac_crit}}
\end{figure}

The emergence of scaling in the vicinity of the critical point is further supported by the 
behavior of the fraction of intact fibers $n$. Figure \ref{fig:intfrac_crit} demonstrates 
that reploting $n$ as a function of the distance from the critical point $E_c-E_0$
an excellent power law is obtained for the zero disorder case
\begin{equation}
 n\sim (E_c-E_0)^{\beta}.
\end{equation}
The value of the exponent is $\beta=1/2$. In the perforation phase $E_0>E_c$ the fraction 
of intact fibers $n$ is identically zero, while in the phase of partial failure 
it has a non zero value $n>0$ going to zero as a power law of the distance from the 
critical point when approaching $E_c$. Due to this behavior, $n$ can be considered as the order parameter 
of the transition and $\beta$ is the order parameter exponent. At finite disorder $W>0$,
the qualitative behavior of the curves is similar to the deflection rate, i.e.\ 
the power law functional form prevails, however, a crossover occurs again between two 
different exponents. Close to failure $n$ approaches zero with a higher exponent $\beta=2/3$, 
however, further from the critical point the exponent takes its zero disorder value 
$\beta=1/2$ (see Fig.\ \ref{fig:intfrac_crit}).

\begin{figure}
\epsfig{bbllx=30,bblly=20,bburx=380,bbury=330, file=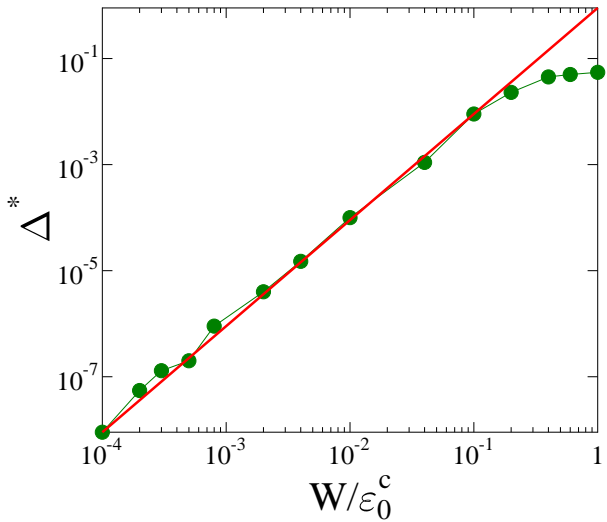, width=8.3cm}
\caption{The value of the crossover point $\Delta^*$ as a function of the degree of disorder 
$W/\varepsilon_0^c$. The straight line represents a power law of exponent 2.0.
\label{fig:crossover}}
\end{figure}

Comparing Figs.\ \ref{fig:deriv},\ref{fig:intfrac_crit} it can be observed that the position of the 
crossover point $\Delta^*$ depends on the degree of disorder: 
As the disorder increases the crossover where the exponents $\gamma$ and $\beta$ switch 
to their higher value, occurs earlier at larger distances from the critical point.
Using the data of the deflection rate in Fig.\ \ref{fig:deriv} we determined the value of the 
crossover point $\Delta^*$ as the position of intersection of two fitted straight lines of exponents 
$1/2$ and $2/3$ in a double logarithmic representation. The results are presented in Fig.\ 
\ref{fig:crossover} where $\Delta^*$ is plotted as a function of $W$. It can be observed 
that approaching the limit of zero disorder $\Delta^*$ goes to zero since the entire curves 
of $n$ and $\delta_m^{\prime}$ are characterized by a single exponent of value $1/2$. 
Increasing the degree of disorder $W$, the value of $\Delta^*$ rapidly grows and levels off
above $W/\varepsilon_0^c\approx 0.15$. The figure also demonstrates that along the increasing regime 
the curve of $\Delta^*(W)$ can be very well approximated by a power law 
\begin{equation}
\Delta^* \sim W^{\xi},
\end{equation}
where the exponent is $\xi=2.0\pm 0.05$.

The saturation of the crossover point $\Delta^*$ implies that beyond a certain value of 
$W$ further increasing the amount of disorder does not have a relevant effect. 
To understand the emergence of this behaviour, the spatial structure 
of damaging has to be analyzed.
At zero disorder perforation occurs such that a crack starts at the bottom of the interface and
proceeds upwards as the impactor moves forward. On the contrary, in the presence of disorder 
$W>0$, fibers break in a spatially random sequence along the interface. Due to the interplay of the strain gradient Eq.\ (\ref{eq:gradient}) and the disorder Eq.\ (\ref{eq:probdens}), a complex 
damage profile emerges along the interface: at a given deflection $\delta$ highly deformed 
fibers at the bottom of the interface are broken with a higher probability forming a crack, 
while the ones on the top have a high chance to be intact. 
The two regimes are separated by a sparse sequence 
of broken and intact fibers which behaves as a process zone ahead of the crack tip. As the 
degree of disorder $W$ is increased the process zone gets wider, and for sufficiently 
large values of $W$ it can span the entire interface. This situation of strong disorder 
occurs if at the deformation
where the top of the interface may get damaged $\varepsilon_1(\delta)>\varepsilon_0^c-W$, the bottom 
of the interface may still be intact $\varepsilon_N(\delta)<\varepsilon_0^c+W$. 
Here $\varepsilon_1$ and $\varepsilon_N$ are the strains of the fibers at the top and bottom
of the bar.
Making use of Eqs.\ (\ref{eq:gradient},\ref{eq:probdens}) and assuming that 
$b\gg \varepsilon_0^c+W$ the condition for strong disorder can be cast into the form
\begin{equation}
W^*\approx \frac{a^2}{2b}\left[\sqrt{1+\frac{4b\varepsilon_0^c}{a^2}}-1\right].
\label{eq:wstar}
\end{equation}
It follows that for $W>W^*$ the disorder is so high that no crack is formed, damage can 
occur anywhere along the interface although it is more probably to break fibers closer to 
the bottom. The results imply that the relevance of disorder in the damaging of the interface 
is determined together by the geometrical layout of the sample $a,b$, by the average strength 
of fibers $\varepsilon_0^c$, and by the width of the strength distribution $W$.
The crossover point $\Delta^*$ depends on $W$ only in the regime of weak disorder $W<W^*$.
The value of $W^*\approx 0.22$ obtained from Eq.\ (\ref{eq:wstar}) has a reasonable agreement
with the numerical findings.

\section{Discussion and conclusions}
We investigated the impact induced fracture of a bar shaped specimen with the 
aim to understand how perforation occurs as the impact energy is gradually
increased. In the modelling approach the bar is represented as two rigid blocks 
glued together by an elastic interface which can undergo damaging as the bar deforms.
The interface is discretized in terms of a bundle of parallel fibers 
which allows for a simple representation of materials' disorder by the random strength 
of fibers. We implemented the loading condition commonly used in the Charpy 
impact test to determine the fracture toughness of materials, i.e.\
a dynamically induced three-point bending test is considered. 

Our analytical and numerical calculations showed that depending on the imparted
energy $E_0$, the outcome of the impact process
can be classified into two states: at low values of the impact energy
the bar suffers only a finite deflection accompanied by damage, resulting in a partial 
failure of the specimen. 
However, exceeding a critical energy value $E_c$ the impact results in global failure
breaking the specimen into two pieces.
The transition between the damaged and perforated phases occurs at a well-defined critical
energy.

In order to characterize how the transition occurs as the impact energy is gradually
increased, we studied the behavior of the deflection rate and of the surviving cross section
when approaching the perforation critical point. Numerical analysis revealed that 
the deflection rate diverges, while the fraction of surviving fibers goes to zero 
as power laws of the distance from the critical point analogous to continuous phase transitions.
The power law behavior holds for any degree of disorder with universal exponents, 
however, at finite disorders 
a crossover occurs between two power law regimes of different exponents: 
further from the critical point the critical exponents coincide with their zero disorder 
counterparts, however, in the vicinity of the critical point both quantities 
are characterized by a higher exponent. The crossover point proved to have a power 
law dependence on the degree of disorder in the range of weak disorder, 
while it remains nearly constant for strongly 
disordered samples. 

The effect of the degree of disorder on the nature of the transition from damage to complete 
breakdown has been studied in various types of systems
\cite{andersen_tricritical_1997,udi_epl_2011,bonamy_failure_2011,ray_epl_2015,biswas_lls_2017,
roy_phasetrans_pre2019,tough_brittle_pre2020}. 
These investigations have revealed that 
below a certain degree of disorder fracture becomes abrupt, so that to obtain power law scaling 
the amount of disorder has to exceed a threshold value. On the contrary, 
it is a unique feature of our system 
that even in the limit of zero disorder the damage - perforation transition occurs analogous 
to continuous phase transitions characterized by critical power laws. The reason of this important 
difference is the loading condition. Our impact loading is assumed to be applied under 
three-point bending conditions so that the stress and strain of fibers is inhomogeneous 
increasing linearly with distance from the impact site. Additionally, the advancing 
impactor gives rise to a loading which is essentially strain controlled, i.e.\ the position of 
the impactor determines the load but in such a way that fiber breakings gradually release the load
on the specimen. The strain gradient in the specimen and the releasing effect of breaking fibers
together stabilize the damage process even in the case of zero disorder. 
For the phase transition nature of the damage-perforation transition the global response of the system,
ensured by the two rigid blocks of the bar, plays an essential role. The main simplification 
of the model is that it does not capture the stress concentration arising at the tip of the 
propagating crack in real materials. However, the model calculations imply that the damage-perforation 
phase transition should occur also in the presence of stress enhancements due to the overall 
bending of the bar.

It can be seen from Eq.\ (\ref{eq:gradient}) that the magnitude of the strain gradient is controlled 
by the geometrical layout of the sample, i.e.\ the gradient is proportional
to the ratio $a/b$ of the sidelengths of the bar. 
Hence, in the limit of $a/b \to 0$, for very long and thin bars all fibers
have nearly the same load. It can be seen from Eq.\ (\ref{eq:wstar}) that in this limit 
the value of $W^*$ separating weak and strong disorders tends to zero so that 
the disorder driven crossover disappears, the system is always in the strong disorder phase.

In our detailed study we considered uniformly distributed breaking thresholds varying the degree 
of disorder through the width of the distribution. Other members
of the universality class of thin-tailed distributions, decreasing rapidly away from the average,
like the Weibull and Gaussian distributions, are expected to give rise qualitatively to the same fracture 
processes \cite{hansen2015fiber}. In particular, we repeated the calculations with the Weibull distribution 
varying the Weibull exponent, while the scale parameter of the distribution was fixed, and obtained  
the same results as in the uniform case. 
However, power law distributed disorder results in a higher degree of complexity. 
Due to the fat tail of the threshold distribution the crossover from 
weak to strong disorder may disappear in spite of the strain gradient. 
Work in this direction is in progress.

\section*{Acknowledgments}
The work is supported by the EFOP-3.6.1-16-2016-00022 project. 
The project is co-financed by the European Union and the European Social Fund.
This research was supported by the National Research, Development and
Innovation Fund of Hungary, financed under the K-16 funding scheme 
Project no.\ K 119967.
The research was financed by the Higher Education Institutional
Excellence Program of the Ministry of Human Capacities in Hungary, 
within the framework of the Energetics thematic program of the University of Debrecen.
The research was financed by the Thematic Excellence Program of the Ministry for
Innovation and Technology in Hungary (ED\_18-1-2019-0028), within the
framework of the Vehicle Industry thematic program of the University of
Debrecen.

\bibliography{/home/feri/papers/statphys_fracture}

\begin{thebibliography}{37}%
\makeatletter
\providecommand \@ifxundefined [1]{%
 \@ifx{#1\undefined}
}%
\providecommand \@ifnum [1]{%
 \ifnum #1\expandafter \@firstoftwo
 \else \expandafter \@secondoftwo
 \fi
}%
\providecommand \@ifx [1]{%
 \ifx #1\expandafter \@firstoftwo
 \else \expandafter \@secondoftwo
 \fi
}%
\providecommand \natexlab [1]{#1}%
\providecommand \enquote  [1]{``#1''}%
\providecommand \bibnamefont  [1]{#1}%
\providecommand \bibfnamefont [1]{#1}%
\providecommand \citenamefont [1]{#1}%
\providecommand \href@noop [0]{\@secondoftwo}%
\providecommand \href [0]{\begingroup \@sanitize@url \@href}%
\providecommand \@href[1]{\@@startlink{#1}\@@href}%
\providecommand \@@href[1]{\endgroup#1\@@endlink}%
\providecommand \@sanitize@url [0]{\catcode `\\12\catcode `\$12\catcode
  `\&12\catcode `\#12\catcode `\^12\catcode `\_12\catcode `\%12\relax}%
\providecommand \@@startlink[1]{}%
\providecommand \@@endlink[0]{}%
\providecommand \url  [0]{\begingroup\@sanitize@url \@url }%
\providecommand \@url [1]{\endgroup\@href {#1}{\urlprefix }}%
\providecommand \urlprefix  [0]{URL }%
\providecommand \Eprint [0]{\href }%
\providecommand \doibase [0]{http://dx.doi.org/}%
\providecommand \selectlanguage [0]{\@gobble}%
\providecommand \bibinfo  [0]{\@secondoftwo}%
\providecommand \bibfield  [0]{\@secondoftwo}%
\providecommand \translation [1]{[#1]}%
\providecommand \BibitemOpen [0]{}%
\providecommand \bibitemStop [0]{}%
\providecommand \bibitemNoStop [0]{.\EOS\space}%
\providecommand \EOS [0]{\spacefactor3000\relax}%
\providecommand \BibitemShut  [1]{\csname bibitem#1\endcsname}%
\let\auto@bib@innerbib\@empty
\bibitem [{\citenamefont {Herrmann}\ and\ \citenamefont
  {Roux}(1990)}]{herrmann_statistical_1990}%
  \BibitemOpen
  \bibinfo {editor} {\bibfnamefont {H.~J.}\ \bibnamefont {Herrmann}}\ and\
  \bibinfo {editor} {\bibfnamefont {S.}~\bibnamefont {Roux}},\ eds.,\
  \href@noop {} {\emph {\bibinfo {title} {Statistical models for the fracture
  of disordered media}}},\ Random materials and processes\ (\bibinfo
  {publisher} {Elsevier},\ \bibinfo {address} {Amsterdam},\ \bibinfo {year}
  {1990})\BibitemShut {NoStop}%
\bibitem [{\citenamefont {Alava}\ \emph {et~al.}(2006)\citenamefont {Alava},
  \citenamefont {Nukala},\ and\ \citenamefont
  {Zapperi}}]{alava_statistical_2006}%
  \BibitemOpen
  \bibfield  {author} {\bibinfo {author} {\bibfnamefont {M.}~\bibnamefont
  {Alava}}, \bibinfo {author} {\bibfnamefont {P.~K.}\ \bibnamefont {Nukala}}, \
  and\ \bibinfo {author} {\bibfnamefont {S.}~\bibnamefont {Zapperi}},\
  }\href@noop {} {\bibfield  {journal} {\bibinfo  {journal} {Adv. Phys.}\
  }\textbf {\bibinfo {volume} {55}},\ \bibinfo {pages} {349–476} (\bibinfo
  {year} {2006})}\BibitemShut {NoStop}%
\bibitem [{\citenamefont {Biswas}\ \emph {et~al.}(2015)\citenamefont {Biswas},
  \citenamefont {Ray},\ and\ \citenamefont
  {Chakrabarti}}]{book_chakrabarti_2015}%
  \BibitemOpen
  \bibfield  {author} {\bibinfo {author} {\bibfnamefont {S.}~\bibnamefont
  {Biswas}}, \bibinfo {author} {\bibfnamefont {P.}~\bibnamefont {Ray}}, \ and\
  \bibinfo {author} {\bibfnamefont {B.~K.}\ \bibnamefont {Chakrabarti}},\
  }\href@noop {} {\emph {\bibinfo {title} {Statistical Physics of Fracture,
  Beakdown, and Earthquake: Effects of Disorder and Heterogeneity}}},\
  Statistical Physics of Fracture and Breakdown\ (\bibinfo  {publisher} {John
  Wiley \& Sons},\ \bibinfo {address} {New York},\ \bibinfo {year}
  {2015})\BibitemShut {NoStop}%
\bibitem [{\citenamefont {Rosti}\ \emph {et~al.}(2009)\citenamefont {Rosti},
  \citenamefont {Illa}, \citenamefont {Koivisto},\ and\ \citenamefont
  {Alava}}]{rosti_crackling_2009}%
  \BibitemOpen
  \bibfield  {author} {\bibinfo {author} {\bibfnamefont {J.}~\bibnamefont
  {Rosti}}, \bibinfo {author} {\bibfnamefont {X.}~\bibnamefont {Illa}},
  \bibinfo {author} {\bibfnamefont {J.}~\bibnamefont {Koivisto}}, \ and\
  \bibinfo {author} {\bibfnamefont {M.~J.}\ \bibnamefont {Alava}},\ }\href@noop
  {} {\bibfield  {journal} {\bibinfo  {journal} {Journal of Physics D: Applied
  Physics}\ }\textbf {\bibinfo {volume} {42}},\ \bibinfo {pages} {214013}
  (\bibinfo {year} {2009})}\BibitemShut {NoStop}%
\bibitem [{\citenamefont {Garcimartin}\ \emph {et~al.}(1997)\citenamefont
  {Garcimartin}, \citenamefont {Guarino}, \citenamefont {Bellon},\ and\
  \citenamefont {Ciliberto}}]{garcimar_statistical_1997}%
  \BibitemOpen
  \bibfield  {author} {\bibinfo {author} {\bibfnamefont {A.}~\bibnamefont
  {Garcimartin}}, \bibinfo {author} {\bibfnamefont {A.}~\bibnamefont
  {Guarino}}, \bibinfo {author} {\bibfnamefont {L.}~\bibnamefont {Bellon}}, \
  and\ \bibinfo {author} {\bibfnamefont {S.}~\bibnamefont {Ciliberto}},\
  }\href@noop {} {\bibfield  {journal} {\bibinfo  {journal} {Phys. Rev. Lett.}\
  }\textbf {\bibinfo {volume} {79}},\ \bibinfo {pages} {3202} (\bibinfo {year}
  {1997})}\BibitemShut {NoStop}%
\bibitem [{\citenamefont {Santucci}\ \emph {et~al.}(2004)\citenamefont
  {Santucci}, \citenamefont {Vanel},\ and\ \citenamefont
  {Ciliberto}}]{santucci_sub-critical_2004}%
  \BibitemOpen
  \bibfield  {author} {\bibinfo {author} {\bibfnamefont {S.}~\bibnamefont
  {Santucci}}, \bibinfo {author} {\bibfnamefont {L.}~\bibnamefont {Vanel}}, \
  and\ \bibinfo {author} {\bibfnamefont {S.}~\bibnamefont {Ciliberto}},\
  }\href@noop {} {\bibfield  {journal} {\bibinfo  {journal} {Phys. Rev. Lett.}\
  }\textbf {\bibinfo {volume} {93}},\ \bibinfo {pages} {095505} (\bibinfo
  {year} {2004})}\BibitemShut {NoStop}%
\bibitem [{\citenamefont {Meinders}\ and\ \citenamefont
  {Vliet}(2008)}]{meinders_scaling_2008}%
  \BibitemOpen
  \bibfield  {author} {\bibinfo {author} {\bibfnamefont {M.~B.~J.}\
  \bibnamefont {Meinders}}\ and\ \bibinfo {author} {\bibfnamefont {T.~v.}\
  \bibnamefont {Vliet}},\ }\href@noop {} {\bibfield  {journal} {\bibinfo
  {journal} {Phys. Rev. E}\ }\textbf {\bibinfo {volume} {77}},\ \bibinfo
  {pages} {036116} (\bibinfo {year} {2008})}\BibitemShut {NoStop}%
\bibitem [{\citenamefont {Deschanel}\ \emph {et~al.}(2009)\citenamefont
  {Deschanel}, \citenamefont {Vanel}, \citenamefont {Godin},\ and\
  \citenamefont {Ciliberto}}]{deschanel_experimental_2009}%
  \BibitemOpen
  \bibfield  {author} {\bibinfo {author} {\bibfnamefont {S.}~\bibnamefont
  {Deschanel}}, \bibinfo {author} {\bibfnamefont {L.}~\bibnamefont {Vanel}},
  \bibinfo {author} {\bibfnamefont {N.}~\bibnamefont {Godin}}, \ and\ \bibinfo
  {author} {\bibfnamefont {S.}~\bibnamefont {Ciliberto}},\ }\href@noop {}
  {\bibfield  {journal} {\bibinfo  {journal} {J. Stat. Mech.: Theor. Exp.}\
  }\textbf {\bibinfo {volume} {2009}},\ \bibinfo {pages} {P01018} (\bibinfo
  {year} {2009})}\BibitemShut {NoStop}%
\bibitem [{\citenamefont {Salje}\ and\ \citenamefont
  {Dahmen}(2014)}]{salje_crackling_2014}%
  \BibitemOpen
  \bibfield  {author} {\bibinfo {author} {\bibfnamefont {E.~K.}\ \bibnamefont
  {Salje}}\ and\ \bibinfo {author} {\bibfnamefont {K.~A.}\ \bibnamefont
  {Dahmen}},\ }\href@noop {} {\bibfield  {journal} {\bibinfo  {journal} {Annual
  Review of Condensed Matter Physics}\ }\textbf {\bibinfo {volume} {5}},\
  \bibinfo {pages} {233} (\bibinfo {year} {2014})}\BibitemShut {NoStop}%
\bibitem [{\citenamefont {Sornette}(2002)}]{sornette_predictability_2002}%
  \BibitemOpen
  \bibfield  {author} {\bibinfo {author} {\bibfnamefont {D.}~\bibnamefont
  {Sornette}},\ }\href@noop {} {\bibfield  {journal} {\bibinfo  {journal}
  {Proc. Natl. Acad. Sci. USA}\ }\textbf {\bibinfo {volume} {99}},\ \bibinfo
  {pages} {2522} (\bibinfo {year} {2002})}\BibitemShut {NoStop}%
\bibitem [{\citenamefont {Nataf}\ \emph {et~al.}(2014)\citenamefont {Nataf},
  \citenamefont {Castillo-Villa}, \citenamefont {Sellappan}, \citenamefont
  {Kriven}, \citenamefont {Vives}, \citenamefont {Planes},\ and\ \citenamefont
  {Salje}}]{nataf_predicting_2014}%
  \BibitemOpen
  \bibfield  {author} {\bibinfo {author} {\bibfnamefont {G.~F.}\ \bibnamefont
  {Nataf}}, \bibinfo {author} {\bibfnamefont {P.~O.}\ \bibnamefont
  {Castillo-Villa}}, \bibinfo {author} {\bibfnamefont {P.}~\bibnamefont
  {Sellappan}}, \bibinfo {author} {\bibfnamefont {W.~M.}\ \bibnamefont
  {Kriven}}, \bibinfo {author} {\bibfnamefont {E.}~\bibnamefont {Vives}},
  \bibinfo {author} {\bibfnamefont {A.}~\bibnamefont {Planes}}, \ and\ \bibinfo
  {author} {\bibfnamefont {E.~K.~H.}\ \bibnamefont {Salje}},\ }\href@noop {}
  {\bibfield  {journal} {\bibinfo  {journal} {Journal of Physics: Condensed
  Matter}\ }\textbf {\bibinfo {volume} {26}},\ \bibinfo {pages} {275401}
  (\bibinfo {year} {2014})}\BibitemShut {NoStop}%
\bibitem [{\citenamefont {Jiang}\ \emph {et~al.}(2017)\citenamefont {Jiang},
  \citenamefont {Liu}, \citenamefont {Main},\ and\ \citenamefont
  {Salje}}]{salje_main_minecollapse_2017}%
  \BibitemOpen
  \bibfield  {author} {\bibinfo {author} {\bibfnamefont {X.}~\bibnamefont
  {Jiang}}, \bibinfo {author} {\bibfnamefont {H.}~\bibnamefont {Liu}}, \bibinfo
  {author} {\bibfnamefont {I.~G.}\ \bibnamefont {Main}}, \ and\ \bibinfo
  {author} {\bibfnamefont {E.~K.~H.}\ \bibnamefont {Salje}},\ }\href@noop {}
  {\bibfield  {journal} {\bibinfo  {journal} {Phys. Rev. E}\ }\textbf {\bibinfo
  {volume} {96}},\ \bibinfo {pages} {023004} (\bibinfo {year}
  {2017})}\BibitemShut {NoStop}%
\bibitem [{\citenamefont {K\'ad\'ar}\ \emph {et~al.}(2020)\citenamefont
  {K\'ad\'ar}, \citenamefont {P\'al},\ and\ \citenamefont
  {Kun}}]{kadar_record_2020}%
  \BibitemOpen
  \bibfield  {author} {\bibinfo {author} {\bibfnamefont {V.}~\bibnamefont
  {K\'ad\'ar}}, \bibinfo {author} {\bibfnamefont {G.}~\bibnamefont {P\'al}}, \
  and\ \bibinfo {author} {\bibfnamefont {F.}~\bibnamefont {Kun}},\ }\href@noop
  {} {\bibfield  {journal} {\bibinfo  {journal} {Scientific Reports}\ }\textbf
  {\bibinfo {volume} {10}},\ \bibinfo {pages} {2508} (\bibinfo {year}
  {2020})}\BibitemShut {NoStop}%
\bibitem [{\citenamefont {Zapperi}\ \emph {et~al.}(1997)\citenamefont
  {Zapperi}, \citenamefont {Ray}, \citenamefont {Stanley},\ and\ \citenamefont
  {Vespignani}}]{zapperi_first-order_1997}%
  \BibitemOpen
  \bibfield  {author} {\bibinfo {author} {\bibfnamefont {S.}~\bibnamefont
  {Zapperi}}, \bibinfo {author} {\bibfnamefont {P.}~\bibnamefont {Ray}},
  \bibinfo {author} {\bibfnamefont {H.~E.}\ \bibnamefont {Stanley}}, \ and\
  \bibinfo {author} {\bibfnamefont {A.}~\bibnamefont {Vespignani}},\
  }\href@noop {} {\bibfield  {journal} {\bibinfo  {journal} {Phys. Rev. Lett.}\
  }\textbf {\bibinfo {volume} {78}},\ \bibinfo {pages} {1408} (\bibinfo {year}
  {1997})}\BibitemShut {NoStop}%
\bibitem [{\citenamefont {Roy}\ \emph {et~al.}(2017)\citenamefont {Roy},
  \citenamefont {Biswas},\ and\ \citenamefont {Ray}}]{biswas_lls_2017}%
  \BibitemOpen
  \bibfield  {author} {\bibinfo {author} {\bibfnamefont {S.}~\bibnamefont
  {Roy}}, \bibinfo {author} {\bibfnamefont {S.}~\bibnamefont {Biswas}}, \ and\
  \bibinfo {author} {\bibfnamefont {P.}~\bibnamefont {Ray}},\ }\href@noop {}
  {\bibfield  {journal} {\bibinfo  {journal} {Phys. Rev. E}\ }\textbf {\bibinfo
  {volume} {96}},\ \bibinfo {pages} {063003} (\bibinfo {year}
  {2017})}\BibitemShut {NoStop}%
\bibitem [{\citenamefont {Buchel}\ and\ \citenamefont
  {Sethna}(1996)}]{sethna_prl1996}%
  \BibitemOpen
  \bibfield  {author} {\bibinfo {author} {\bibfnamefont {A.}~\bibnamefont
  {Buchel}}\ and\ \bibinfo {author} {\bibfnamefont {J.~P.}\ \bibnamefont
  {Sethna}},\ }\href@noop {} {\bibfield  {journal} {\bibinfo  {journal} {Phys.
  Rev. Lett.}\ }\textbf {\bibinfo {volume} {77}},\ \bibinfo {pages} {1520}
  (\bibinfo {year} {1996})}\BibitemShut {NoStop}%
\bibitem [{\citenamefont {Zapperi}\ \emph {et~al.}(1999)\citenamefont
  {Zapperi}, \citenamefont {Ray}, \citenamefont {Stanley},\ and\ \citenamefont
  {Vespignani}}]{zapperi_analysis_1999}%
  \BibitemOpen
  \bibfield  {author} {\bibinfo {author} {\bibfnamefont {S.}~\bibnamefont
  {Zapperi}}, \bibinfo {author} {\bibfnamefont {P.}~\bibnamefont {Ray}},
  \bibinfo {author} {\bibfnamefont {H.~E.}\ \bibnamefont {Stanley}}, \ and\
  \bibinfo {author} {\bibfnamefont {A.}~\bibnamefont {Vespignani}},\
  }\href@noop {} {\bibfield  {journal} {\bibinfo  {journal} {Physica A}\
  }\textbf {\bibinfo {volume} {270}},\ \bibinfo {pages} {57} (\bibinfo {year}
  {1999})}\BibitemShut {NoStop}%
\bibitem [{\citenamefont {Kun}\ and\ \citenamefont
  {Herrmann}(2000)}]{kun_damage_2000}%
  \BibitemOpen
  \bibfield  {author} {\bibinfo {author} {\bibfnamefont {F.}~\bibnamefont
  {Kun}}\ and\ \bibinfo {author} {\bibfnamefont {H.~J.}\ \bibnamefont
  {Herrmann}},\ }\href@noop {} {\bibfield  {journal} {\bibinfo  {journal} {J.
  Mat. Sci.}\ }\textbf {\bibinfo {volume} {35}},\ \bibinfo {pages} {4685}
  (\bibinfo {year} {2000})}\BibitemShut {NoStop}%
\bibitem [{\citenamefont {Sornette}\ and\ \citenamefont
  {Andersen}(1998)}]{sornette_scaling_1998}%
  \BibitemOpen
  \bibfield  {author} {\bibinfo {author} {\bibfnamefont {D.}~\bibnamefont
  {Sornette}}\ and\ \bibinfo {author} {\bibfnamefont {J.}~\bibnamefont
  {Andersen}},\ }\href@noop {} {\bibfield  {journal} {\bibinfo  {journal} {Eur.
  Phys. J. B}\ }\textbf {\bibinfo {volume} {1}},\ \bibinfo {pages} {353}
  (\bibinfo {year} {1998})}\BibitemShut {NoStop}%
\bibitem [{\citenamefont {Moreno}\ \emph {et~al.}(2000)\citenamefont {Moreno},
  \citenamefont {Gomez},\ and\ \citenamefont {Pacheco}}]{moreno_fracture_2000}%
  \BibitemOpen
  \bibfield  {author} {\bibinfo {author} {\bibfnamefont {Y.}~\bibnamefont
  {Moreno}}, \bibinfo {author} {\bibfnamefont {J.~B.}\ \bibnamefont {Gomez}}, \
  and\ \bibinfo {author} {\bibfnamefont {A.~F.}\ \bibnamefont {Pacheco}},\
  }\href@noop {} {\bibfield  {journal} {\bibinfo  {journal} {Phys. Rev. Lett.}\
  }\textbf {\bibinfo {volume} {85}},\ \bibinfo {pages} {2865} (\bibinfo {year}
  {2000})}\BibitemShut {NoStop}%
\bibitem [{\citenamefont {Caldarelli}\ \emph {et~al.}(1996)\citenamefont
  {Caldarelli}, \citenamefont {Tolla},\ and\ \citenamefont
  {Petri}}]{caldarelli_self_organization_1996}%
  \BibitemOpen
  \bibfield  {author} {\bibinfo {author} {\bibfnamefont {G.}~\bibnamefont
  {Caldarelli}}, \bibinfo {author} {\bibfnamefont {F.~D.~D.}\ \bibnamefont
  {Tolla}}, \ and\ \bibinfo {author} {\bibfnamefont {A.}~\bibnamefont
  {Petri}},\ }\href@noop {} {\bibfield  {journal} {\bibinfo  {journal} {Phys.
  Rev. Lett.}\ }\textbf {\bibinfo {volume} {77}},\ \bibinfo {pages} {2503}
  (\bibinfo {year} {1996})}\BibitemShut {NoStop}%
\bibitem [{\citenamefont {Bonamy}\ \emph {et~al.}(2008)\citenamefont {Bonamy},
  \citenamefont {Santucci},\ and\ \citenamefont
  {Ponson}}]{bonamy_crackling_prl2008}%
  \BibitemOpen
  \bibfield  {author} {\bibinfo {author} {\bibfnamefont {D.}~\bibnamefont
  {Bonamy}}, \bibinfo {author} {\bibfnamefont {S.}~\bibnamefont {Santucci}}, \
  and\ \bibinfo {author} {\bibfnamefont {L.}~\bibnamefont {Ponson}},\
  }\href@noop {} {\bibfield  {journal} {\bibinfo  {journal} {Phys. Rev. Lett.}\
  }\textbf {\bibinfo {volume} {101}},\ \bibinfo {pages} {045501} (\bibinfo
  {year} {2008})}\BibitemShut {NoStop}%
\bibitem [{\citenamefont {Roy}\ and\ \citenamefont
  {Manna}(2019)}]{roy_phasetrans_pre2019}%
  \BibitemOpen
  \bibfield  {author} {\bibinfo {author} {\bibfnamefont {C.}~\bibnamefont
  {Roy}}\ and\ \bibinfo {author} {\bibfnamefont {S.~S.}\ \bibnamefont
  {Manna}},\ }\href@noop {} {\bibfield  {journal} {\bibinfo  {journal} {Phys.
  Rev. E}\ }\textbf {\bibinfo {volume} {100}},\ \bibinfo {pages} {012107}
  (\bibinfo {year} {2019})}\BibitemShut {NoStop}%
\bibitem [{\citenamefont {Picallo}\ \emph {et~al.}(2010)\citenamefont
  {Picallo}, \citenamefont {L\'opez}, \citenamefont {Zapperi},\ and\
  \citenamefont {Alava}}]{picallo_brittle_2010}%
  \BibitemOpen
  \bibfield  {author} {\bibinfo {author} {\bibfnamefont {C.~B.}\ \bibnamefont
  {Picallo}}, \bibinfo {author} {\bibfnamefont {J.~M.}\ \bibnamefont
  {L\'opez}}, \bibinfo {author} {\bibfnamefont {S.}~\bibnamefont {Zapperi}}, \
  and\ \bibinfo {author} {\bibfnamefont {M.~J.}\ \bibnamefont {Alava}},\
  }\href@noop {} {\bibfield  {journal} {\bibinfo  {journal} {Phys. Rev. Lett.}\
  }\textbf {\bibinfo {volume} {105}},\ \bibinfo {pages} {155502} (\bibinfo
  {year} {2010})}\BibitemShut {NoStop}%
\bibitem [{\citenamefont {Karpas}\ and\ \citenamefont
  {Kun}(2011)}]{udi_epl_2011}%
  \BibitemOpen
  \bibfield  {author} {\bibinfo {author} {\bibfnamefont {E.}~\bibnamefont
  {Karpas}}\ and\ \bibinfo {author} {\bibfnamefont {F.}~\bibnamefont {Kun}},\
  }\href@noop {} {\bibfield  {journal} {\bibinfo  {journal} {Europhys. Lett.}\
  }\textbf {\bibinfo {volume} {95}},\ \bibinfo {pages} {16004} (\bibinfo {year}
  {2011})}\BibitemShut {NoStop}%
\bibitem [{\citenamefont {Danku}\ and\ \citenamefont
  {Kun}(2016)}]{danku_disorder_2016}%
  \BibitemOpen
  \bibfield  {author} {\bibinfo {author} {\bibfnamefont {Z.}~\bibnamefont
  {Danku}}\ and\ \bibinfo {author} {\bibfnamefont {F.}~\bibnamefont {Kun}},\
  }\href@noop {} {\bibfield  {journal} {\bibinfo  {journal} {J. Stat. Mech.:
  Theor. Exp.}\ }\textbf {\bibinfo {volume} {2016}},\ \bibinfo {pages} {073211}
  (\bibinfo {year} {2016})}\BibitemShut {NoStop}%
\bibitem [{\citenamefont {Roy}\ \emph {et~al.}(2015)\citenamefont {Roy},
  \citenamefont {Kundu},\ and\ \citenamefont
  {Manna}}]{manna_PhysRevE.91.032103}%
  \BibitemOpen
  \bibfield  {author} {\bibinfo {author} {\bibfnamefont {C.}~\bibnamefont
  {Roy}}, \bibinfo {author} {\bibfnamefont {S.}~\bibnamefont {Kundu}}, \ and\
  \bibinfo {author} {\bibfnamefont {S.~S.}\ \bibnamefont {Manna}},\ }\href@noop
  {} {\bibfield  {journal} {\bibinfo  {journal} {Phys. Rev. E}\ }\textbf
  {\bibinfo {volume} {91}},\ \bibinfo {pages} {032103} (\bibinfo {year}
  {2015})}\BibitemShut {NoStop}%
\bibitem [{\citenamefont {Ramos}\ \emph {et~al.}(2013)\citenamefont {Ramos},
  \citenamefont {Cortet}, \citenamefont {Ciliberto},\ and\ \citenamefont
  {Vanel}}]{ramos_prl_2013}%
  \BibitemOpen
  \bibfield  {author} {\bibinfo {author} {\bibfnamefont {O.}~\bibnamefont
  {Ramos}}, \bibinfo {author} {\bibfnamefont {P.-P.}\ \bibnamefont {Cortet}},
  \bibinfo {author} {\bibfnamefont {S.}~\bibnamefont {Ciliberto}}, \ and\
  \bibinfo {author} {\bibfnamefont {L.}~\bibnamefont {Vanel}},\ }\href@noop {}
  {\bibfield  {journal} {\bibinfo  {journal} {Phys. Rev. Lett.}\ }\textbf
  {\bibinfo {volume} {110}},\ \bibinfo {pages} {165506} (\bibinfo {year}
  {2013})}\BibitemShut {NoStop}%
\bibitem [{\citenamefont {Roy}\ and\ \citenamefont {Ray}(2015)}]{ray_epl_2015}%
  \BibitemOpen
  \bibfield  {author} {\bibinfo {author} {\bibfnamefont {S.}~\bibnamefont
  {Roy}}\ and\ \bibinfo {author} {\bibfnamefont {P.}~\bibnamefont {Ray}},\
  }\href@noop {} {\bibfield  {journal} {\bibinfo  {journal} {Europhys. Lett.}\
  }\textbf {\bibinfo {volume} {112}},\ \bibinfo {pages} {26004} (\bibinfo
  {year} {2015})}\BibitemShut {NoStop}%
\bibitem [{\citenamefont {Siewert}\ and\ \citenamefont
  {Manahan}(2000)}]{siewert_pendulum_2000}%
  \BibitemOpen
  \bibfield  {author} {\bibinfo {author} {\bibfnamefont {T.~A.}\ \bibnamefont
  {Siewert}}\ and\ \bibinfo {author} {\bibfnamefont {M.~P.}\ \bibnamefont
  {Manahan}},\ }\href@noop {} {\emph {\bibinfo {title} {Pendulum Impact
  Testing: A Century of Progress}}}\ (\bibinfo  {publisher} {{ASTM}
  International},\ \bibinfo {year} {2000})\BibitemShut {NoStop}%
\bibitem [{\citenamefont {Francois}\ and\ \citenamefont
  {Pineau}(2002)}]{francois_charpy_2002}%
  \BibitemOpen
  \bibfield  {author} {\bibinfo {author} {\bibfnamefont {D.}~\bibnamefont
  {Francois}}\ and\ \bibinfo {author} {\bibfnamefont {A.}~\bibnamefont
  {Pineau}},\ }\href@noop {} {\emph {\bibinfo {title} {From Charpy to Present
  Impact Testing}}}\ (\bibinfo  {publisher} {Elsevier, Amsterdam},\ \bibinfo
  {year} {2002})\BibitemShut {NoStop}%
\bibitem [{\citenamefont {Kun}\ \emph {et~al.}(2004)\citenamefont {Kun},
  \citenamefont {Lenkey}, \citenamefont {Tak\'acs},\ and\ \citenamefont
  {Beke}}]{kun_structure_2004}%
  \BibitemOpen
  \bibfield  {author} {\bibinfo {author} {\bibfnamefont {F.}~\bibnamefont
  {Kun}}, \bibinfo {author} {\bibfnamefont {G.~B.}\ \bibnamefont {Lenkey}},
  \bibinfo {author} {\bibfnamefont {N.}~\bibnamefont {Tak\'acs}}, \ and\
  \bibinfo {author} {\bibfnamefont {D.~L.}\ \bibnamefont {Beke}},\ }\href@noop
  {} {\bibfield  {journal} {\bibinfo  {journal} {Phys. Rev. Lett.}\ }\textbf
  {\bibinfo {volume} {93}},\ \bibinfo {pages} {227204} (\bibinfo {year}
  {2004})}\BibitemShut {NoStop}%
\bibitem [{\citenamefont {Danku}\ \emph {et~al.}(2015)\citenamefont {Danku},
  \citenamefont {Lenkey},\ and\ \citenamefont {Kun}}]{danku_apl_2015}%
  \BibitemOpen
  \bibfield  {author} {\bibinfo {author} {\bibfnamefont {Z.}~\bibnamefont
  {Danku}}, \bibinfo {author} {\bibfnamefont {G.~B.}\ \bibnamefont {Lenkey}}, \
  and\ \bibinfo {author} {\bibfnamefont {F.}~\bibnamefont {Kun}},\ }\href@noop
  {} {\bibfield  {journal} {\bibinfo  {journal} {Applied Physics Letters}\
  }\textbf {\bibinfo {volume} {106}},\ \bibinfo {pages} {064102} (\bibinfo
  {year} {2015})}\BibitemShut {NoStop}%
\bibitem [{\citenamefont {Andersen}\ \emph {et~al.}(1997)\citenamefont
  {Andersen}, \citenamefont {Sornette},\ and\ \citenamefont
  {Leung}}]{andersen_tricritical_1997}%
  \BibitemOpen
  \bibfield  {author} {\bibinfo {author} {\bibfnamefont {J.~V.}\ \bibnamefont
  {Andersen}}, \bibinfo {author} {\bibfnamefont {D.}~\bibnamefont {Sornette}},
  \ and\ \bibinfo {author} {\bibfnamefont {K.}~\bibnamefont {Leung}},\
  }\href@noop {} {\bibfield  {journal} {\bibinfo  {journal} {Phys. Rev. Lett.}\
  }\textbf {\bibinfo {volume} {78}},\ \bibinfo {pages} {2140–2143} (\bibinfo
  {year} {1997})}\BibitemShut {NoStop}%
\bibitem [{\citenamefont {Bonamy}\ and\ \citenamefont
  {Bouchaud}(2011)}]{bonamy_failure_2011}%
  \BibitemOpen
  \bibfield  {author} {\bibinfo {author} {\bibfnamefont {D.}~\bibnamefont
  {Bonamy}}\ and\ \bibinfo {author} {\bibfnamefont {E.}~\bibnamefont
  {Bouchaud}},\ }\href@noop {} {\bibfield  {journal} {\bibinfo  {journal}
  {Physics Reports}\ }\textbf {\bibinfo {volume} {498}},\ \bibinfo {pages} {1}
  (\bibinfo {year} {2011})}\BibitemShut {NoStop}%
\bibitem [{\citenamefont {Kachhwah}\ and\ \citenamefont
  {Mahesh}(2020)}]{tough_brittle_pre2020}%
  \BibitemOpen
  \bibfield  {author} {\bibinfo {author} {\bibfnamefont {U.~S.}\ \bibnamefont
  {Kachhwah}}\ and\ \bibinfo {author} {\bibfnamefont {S.}~\bibnamefont
  {Mahesh}},\ }\href@noop {} {\bibfield  {journal} {\bibinfo  {journal} {Phys.
  Rev. E}\ }\textbf {\bibinfo {volume} {101}},\ \bibinfo {pages} {063002}
  (\bibinfo {year} {2020})}\BibitemShut {NoStop}%
\bibitem [{\citenamefont {Hansen}\ \emph {et~al.}(2015)\citenamefont {Hansen},
  \citenamefont {Hemmer},\ and\ \citenamefont {Pradhan}}]{hansen2015fiber}%
  \BibitemOpen
  \bibfield  {author} {\bibinfo {author} {\bibfnamefont {A.}~\bibnamefont
  {Hansen}}, \bibinfo {author} {\bibfnamefont {P.}~\bibnamefont {Hemmer}}, \
  and\ \bibinfo {author} {\bibfnamefont {S.}~\bibnamefont {Pradhan}},\
  }\href@noop {} {\emph {\bibinfo {title} {The Fiber Bundle Model: Modeling
  Failure in Materials}}},\ Statistical Physics of Fracture and Breakdown\
  (\bibinfo  {publisher} {Wiley},\ \bibinfo {year} {2015})\BibitemShut
  {NoStop}%
\end{thebibliography}%

\end{document}